\documentclass[twocolumn,dvipsnames]{aastex62}
\usepackage{amsmath,amstext}
\usepackage{apjfonts}
\usepackage[figure,figure*]{hypcap}
\usepackage{ulem}
\usepackage{xspace}
\usepackage{lineno}
\usepackage{xfrac}
\usepackage{graphicx}
\usepackage{natbib}
\usepackage{multimedia}
\usepackage{amssymb}
\usepackage{amsmath}
\usepackage{array,booktabs}
\usepackage{mathptmx}
\usepackage{booktabs}
\usepackage{hyperref}
\usepackage{float}
\usepackage{todonotes}
\usepackage{multirow}
\newcommand{\rej}{\,{r_{\rm ej}}}
\newcommand{\vej}{\,{v_{\rm ej}}}
\newcommand{\teng}{\,{t_{\rm eng}}}
\newcommand{\reji}{\,{r_{\rm ej,0}}}
\newcommand{\erg}{\,{{\rm erg}}}

\newcommand{\msun}{\,{M_{\odot}}}

\newcommand{\s}{\,{{\rm s}}}
\newcommand{\days}{\,{{\rm d}}}

\newcommand{\cm}{\,{{\rm cm}}}
\newcommand{\km}{\,{{\rm km}}}
\newcommand{\Lj}{\,{L_{\rm j}}}
\newcommand{\Eth}{\,{E_{\rm th}}}
\newcommand{\rph}{\,{R_{\rm ph}}}
\newcommand{\rtr}{\,{R_{\rm tr}}}
\newcommand{\vemit}{\,{\beta_{\rm emit}}}

\shorttitle{Illuminating the Early Bumps in Superluminous Supernovae}
\shortauthors{Gottlieb \& Metzger}

\begin{document}

\title{Late Jets, Early Sparks: Illuminating the Premaximum Bumps in Superluminous Supernovae}

    \author[0000-0003-3115-2456]{Ore Gottlieb}
	\email{ogottlieb@flatironinstitute.org}
    \affiliation{Center for Computational Astrophysics, Flatiron Institute, 162 5th Avenue, New York, NY 10010, USA}
    \affil{Department of Physics and Columbia Astrophysics Laboratory, Columbia University, Pupin Hall, New York, NY 10027, USA}
    
    \author[0000-0002-3635-5677]{Brian D. Metzger}
    \affiliation{Department of Physics and Columbia Astrophysics Laboratory, Columbia University, Pupin Hall, New York, NY 10027, USA}
    \affiliation{Center for Computational Astrophysics, Flatiron Institute, 162 5th Avenue, New York, NY 10010, USA}

\begin{abstract}
Superluminous supernovae (SLSNe) radiate $\gtrsim 10-100$ times more energy than ordinary stellar explosions, implicating a novel power source behind these enigmatic events. One frequently discussed source, particularly for hydrogen-poor (Type I) SLSNe, is a central engine such as a millisecond magnetar or accreting black hole. Both black hole and magnetar engines are expected to channel a fraction of their luminosity into a collimated relativistic jet. Using 3D relativistic hydrodynamical simulations, we explore the interaction of a relativistic jet, endowed with a luminosity $L_{\rm j}\approx10^{45.5}\,{\rm erg\,s^{-1}}$ and duration $t_{\rm eng} \approx 10\,{\rm days}$ compatible with those needed to power SLSNe, launched into the envelope of the exploding star. The jet successfully breaks through the expanding ejecta, and its shocked cocoon powers ultraviolet/optical emission lasting several days after the explosion and reaching a peak luminosity $\gtrsim 10^{44}\,{\rm erg\,s^{-1}} $, corresponding to a sizable fraction of $L_{\rm j}$. This high radiative efficiency is the result of the modest adiabatic losses the cocoon experiences owing to the low optical depths of the enlarged ejecta at these late times, e.g., compared to the more compact stars in gamma-ray bursts. The luminosity and temperature of the cocoon emission match those of the ``bumps'' in SLSN light curves observed weeks prior to the optical maximum in many SLSNe. Confirmation of jet breakout signatures by future observations (e.g., days-long to weeks-long internal X-ray emission from the jet for on-axis observers, spectroscopy confirming large photosphere velocities $v/c \gtrsim 0.1$, or detection of a radio afterglow) would offer strong evidence for central engines powering SLSNe.
\end{abstract}

\section{Introduction}

The collapse of a rotating massive star giving birth to a magnetized spinning compact object -- either a Kerr black hole or a neutron star -- offers a compelling scenario for generating some of the most energetic transients in the Universe, including gamma-ray bursts (GRBs) and superluminous supernovae (SLSNe). While most GRBs with long durations are accompanied by stripped-envelope SNe with abnormally large kinetic energies (``broad-lined'' Type Ibc SNe; e.g., \citealt{Stanek+03,Woosley&Bloom06,Liu&Modjaz16}), SLSNe are instead distinguished by their large {\it radiated} energies, which can exceed those of ordinary stellar explosions by a factor of $\gtrsim 10-100$ (e.g., \citealt{GalYam18,Nicholl21} for reviews). Type I (i.e., hydrogen-poor) SLSNe, in particular, may be associated with the birth of a millisecond (ms) neutron star with a strong magnetic field (``ms magnetar''; e.g., \citealt{Kasen2010,Woosley10,Nicholl+17d,Vurm&Metzger21}). The magnetized wind that accompanies the first tens of seconds in the life of a ms magnetar can feed a relativistic jet with the necessary luminosity and baryon loading (Lorentz factor) to power a GRB (e.g., \citealt{Wheeler+00,Thompson+04,Bucciantini+08,Metzger+11}), though the accretion-powered jet of a Kerr black hole remains a more commonly invoked scenario (e.g., \citealt{Woosley93,MacFadyen&Woosley99,Gottlieb2023a}).

\begin{figure}
        \centering
        {\includegraphics[width=3.5in,trim={0cm 0cm 0cm 0cm}]{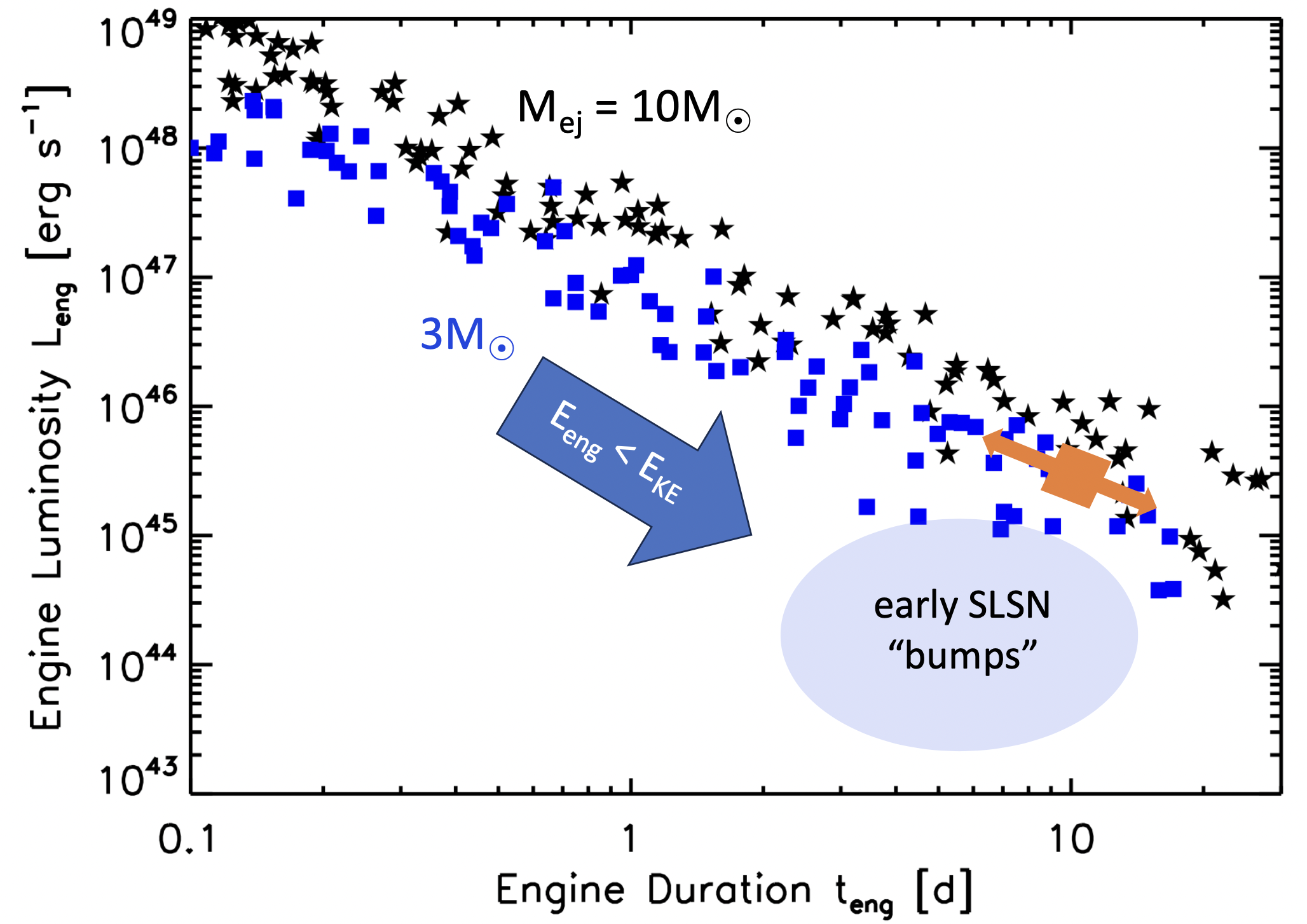}}
        \caption{Required engine peak luminosity, $L_{\rm eng}$ and duration at this luminosity, $t_{\rm eng}$ (e.g., magnetar spin-down luminosity and timescale, respectively), to explain the observed peak luminosities, $L_{\rm pk}$, and rise times, $t_{\rm pk}$, of a large sample of Type I SLSN light curves \citep{Gomez+24}, for two different assumptions about the ejecta mass, $M_{\rm ej} = 10M_{\odot}$ (black stars) and $3M_{\odot}$ (blue squares), bracketing the typical range inferred for Type I SLSNe \citep{Nicholl+17d}. For comparison, we show the approximate range of timescales and luminosities of the early premaximum bumps observed in the light curves of Type I SLSNe (shaded blue ellipse; see e.g. \citealt{Nicholl&Smartt16}). A multiday-lasting jet from the same central engine responsible for powering the bulk of the SLSNe is thus capable of powering the early bumps, assuming that the jet luminosity, $ \Lj $, is a fraction $\sim 1-10\%$ of $L_{\rm eng}$. The assumed jet properties for the numerical simulation presented in this paper are illustrated with an orange square, while bidirectional arrows indicate the degeneracy in engine properties $L_{\rm eng} \propto t_{\rm eng}^{-1}$ that result in the same hydrodynamic evolution for fixed ejecta properties. 
        {\it Details:} For each event, we estimate the ejecta kinetic energy, $E_{\rm KE} = M_{\rm ej}\vej^{2}/2$, by equating the observed SN rise-timescale to the diffusion timescale for a centrally-concentrated energy source, $t_{\rm diff} \simeq 0.4(\kappa M_{\rm ej}/\vej c)^{1/2}$ \citep{Khatami&Kasen19}. Assuming that the explosion energy $E_{\rm KE}$ comes exclusively from the engine, i.e. $E_{\rm eng} = L_{\rm eng}t_{\rm eng} = E_{\rm KE}$, we estimate the required engine timescale assuming that the engine luminosity decays similarly to magnetic-dipole spin-down $L_{\rm pk} = (E_{\rm eng}/t_{\rm eng})(1+t_{\rm pk}/t_{\rm eng})^{-2}$, for $t_{\rm eng} < t_{\rm pk}$.  If the engine contributes only a fraction of the explosion energy $E_{\rm KE}$, then a greater engine duration is needed to explain the observed luminosities and points will shift towards the lower right part of the diagram as indicated by the arrow.}
        \label{fig:engineproperties}
    \end{figure}

What distinguishes the central engines of GRBs versus those that power SLSNe? Although the total energetics of both systems can be comparable, their main distinction is the {\it duration} of the central engine activity \citep{Metzger+15}. While GRB engines typically emit a sizable fraction of their energy \citep{Panaitescu&Kumar02} over seconds to minutes (comparable to the duration of the gamma-ray emission; e.g., \citealt{Bromberg+12}), the engines of SLSNe must instead heat the SN ejecta continuously from within over much longer timescales of days to weeks. This heating being substantially delayed after the time of the stellar explosion is crucial to explaining the high radiative output of SLSNe. A delay ensures that less deposited energy is ``wasted'' through adiabatic losses by occurring at later times after the optical depth through the expanding ejecta has dropped. Fig.~\ref{fig:engineproperties} shows estimates of the engine duration and luminosity required to explain the rise times and peak luminosities of a large sample of Type I SLSNe \citep{Gomez+24}.

In magnetar scenarios, the engine duration is typically related to the magnetic dipole spin-down timescale $t_{\rm sd} \propto B^{2}$ which varies considerably with the surface magnetic field strength ($B \sim 10^{14}$ G to power SLSNe vs. $B \gtrsim 3\times 10^{15}$ G to power a typical long GRB). Even at fixed $B$, differences in the magnetar's mass accretion rate \citep{Metzger+18b} or the inclination angle of the magnetic dipole relative to the spin axis \citep{Margalit+18b} can lead to large differences in the jet power. So-called ``ultralong'' GRBs (e.g., \citealt{Levan+14}) require engine durations of hours, commensurate with their longer gamma-ray emission; these events may represent an intermediate case in which the SN's luminosity is boosted only modestly above the baseline level set by $^{56}$Ni decay \citep{Greiner+15,Metzger+15}, the latter dominating in ordinary GRB SNe (e.g., \citealt{Woosley+02}).

While a variety of indirect evidence supports a compact object residing at the center of at least some Type I SLSNe (e.g., \citealt{Lunnan+15,Eftekhari+19,Nicholl+19,Vurm&Metzger21,Omand&Jerkstrand23}), a central engine is not the only plausible power source for these events.  A commonly invoked alternative is shock interaction between the fast SN ejecta and a compact wind or other circumstellar medium (CSM) that surrounds the progenitor star at the time of explosion (e.g., \citealt{Chevalier&Irwin11,Chatzopoulos+13,Lunnan+18,Fraser20,Aguilera-Dena+20,Stevance&Eldridge21}). Given the challenges in cleanly distinguishing engine from CSM interaction models, alternative tests of the engine hypothesis are sorely needed. Gamma-ray emission from a magnetar nebula on timescales of months to years after the ejecta becomes optically-thin offers one compelling test (e.g., \citealt{Murase+15,Vurm&Metzger21,Murase+21}); however, the low predicted gamma-ray flux at these late times makes detection challenging in all but the closest explosions \citep{Acharyya+23}. A more promising test is one that probes the central engine at earlier times closer to its maximal power. 

If a rotating compact object powers both GRBs and SLSNe, it is natural to ask whether SLSNe should also be accompanied by ultrarelativistic jets \citep[see][for a review of the role of jets in CCSNe]{Soker2022}. While the luminosities of SLSN engines are lower than those of GRB jets, the density of the external medium through which a putative SLSN jet must propagate is also lower. This is due to the larger ejecta radius $R_{\rm ej} \approx \vej t \sim 10^{13}-10^{14}$ cm at such late times $t\sim$ days$-$weeks after the explosion. Here, $\vej \approx 5000$ km s$^{-1}$ is the speed of the SN ejecta, which for typical stripped-star progenitor radii $R_{\star} \sim 10^{11}$ cm, achieves homologous expansion on a timescale $\gtrsim R_{\star}/\vej \sim 10^{2}$ s. Analytic estimates show that even jet luminosities $\Lj \simeq 10^{45}$ erg s$^{-1}$, much lower than for GRBs but comparable to those required of many Type I SLSNe engines (Fig.~\ref{fig:engineproperties}), are sufficient for jet breakout through the homologously expanding stellar ejecta (e.g., \citealt{Quataert&Kasen12,Margalit+18b,Gottlieb&Nakar22}). In particular, for a jet of half-opening angle $\theta_j$ and luminosity $\Lj$, breakout can occur on a timescale \citep[Eq. 45 in][]{Gottlieb&Nakar22}:
\begin{equation}\label{eq:tbo}
    t_{\rm bo} \approx 5\,\frac{E_{\rm KE}}{3\times 10^{51}\,\erg}\frac{10^{45}\,\erg\,\s^{-1}}{\Lj}\left(\frac{\theta_j}{10^\circ}\right)^4\,\days\,,
\end{equation}
where the prefactor assumes a non-magnetized jet propagation, as expected for extended-duration weak jets, and other quantities including the ejecta kinetic energy $E_{\rm KE}$ are normalized to characteristic values (Fig.~\ref{fig:engineproperties}). The jet breakout times of days indicated by Eq.~\eqref{eq:tbo} are notably similar to the typical engine durations (e.g., magnetar spin-down time) required to power the bulk of the SLSN emission.

Do observations rule out jets in SLSNe? With one interesting exception (\citealt{Levan+13}), no X-ray emission has been detected to faint limits from a sample of 26 Type I SLSNe on timescales of months to years after the explosion (e.g., \citealt{Margutti+18}). However, this is not necessarily constraining because the spectral energy distribution of any (internal) jetted emission is uncertain; and even if the jet efficiently powers X-rays, they may only be detectable at early times and for observers within the narrow opening angle of the jet. In principle, radio afterglow emission from the jet as it interacts with surrounding CSM may be detectable at late times even for slightly off-axis observers; however, again (with one or two interesting exceptions; \citealt{Eftekhari+19,Margutti+23}\footnote{In PTF10hgi \citep{Eftekhari+19,Mondal+20}, radio emission detected starting 7.5 years after the explosion may originate from a magnetar nebula (e.g., \citealt{Margalit&Metzger18,Omand+18}), while the late radio emission in SN 2017ens \citep{Margutti+23} originates from the SN ejecta colliding with CSM.}) no radio detections of Type I SLSNe have been made (e.g., \citealt{Coppejans+18,Eftekhari+21,Hatsukade+21}). As we shall discuss, the resulting constraints on the jet power are sensitive to the observer viewing angle, CSM density, and uncertain microphysical parameters of the shock. 

A cleaner signature of a successful jet in Type I SLSNe may come at earlier times, on timescales comparable to the engine duration of days to weeks \citep{Margalit+18b}. In ordinary GRB jets, the compact size of the shocked stellar debris (the so-called ``cocoon''; e.g., \citealt{Gottlieb2018a}) results in most of the cocoon radiation being initially trapped and only subsequently released after many expansion times. By contrast, for the more sustained jet-ejecta interaction expected for longer-lived SLSN jets interacting with substantially larger ejecta, the lower optical depths (smaller trapping radius in the ejecta frame) will alter the observed cocoon light curve, allowing it to track energy input from the engine more faithfully.  

A suggestive hint comes from the early maxima (``bumps'') of luminosity $\sim 10^{43}-10^{45}$ erg s$^{-1}$ observed in some Type I SLSNe optical light curves on timescales of days to a week after the explosion, well before the main SN peak \citep{Leloudas+12,Nicholl+15b}. Although such premaximum bumps are not detected in all events due to observational limitations, \citet{Nicholl&Smartt16} show that they are likely common features in SLSNe. Subsequent studies have shown that early maxima are not present in all SLSNe, and that their properties are heterogeneous, with a wide range of rise and decline times (e.g., \citealt{Angus+19}). Qualitatively similar ``double peaked'' light curves are seen in a fraction $\sim 3-9\%$ of ordinary-luminosity stripped-envelope SNe (e.g., \citealt{Das+23} and references therein), which however are frequently attributed to shock interaction between the SN ejecta and CSM from pre-explosion mass-loss.

It may not be a coincidence that the timescales of the observed bumps coincide both with the first light from the jet breakout [Eq.~\eqref{eq:tbo}] and the engine durations needed to power SLSN emission. The luminosities of the bumps can also be accommodated within the energy budget of the engine (light blue cloud in Fig.~\ref{fig:engineproperties}). \citet{Kasen+16} proposed that this early emission phase is powered by ejecta reheating due to a delayed spherical shock breakout through the ejecta, distinct from the earlier SN shock breakout and driven by the pressure of a pulsar-like nebula inflated inside the ejecta by the magnetar wind (see also \citealt{Chen+16,Suzuki&Maeda16,Moriya2022}).

In this \emph{Letter}, we investigate whether this early emission phase can instead be powered by a collimated jet$-$from an engine with similar properties needed to power the full SLSN light curve at later times$-$that successfully breaks out of the SN ejecta. In \S\ref{sec:numerical_setup}, we establish a 3D numerical simulation to track the jet propagation inside and outside the ejecta over several weeks \citep[see also][for 2D simulations of jet-SN ejecta interaction]{Ramirez-Ruiz2010,DeColle2022}. In \S\ref{sec:numerical_emission}, we outline the post-processing calculations to estimate the resulting emission. We present our findings in \S\ref{sec:results}, demonstrating their consistency with observations of the early bumps in Type I SLSNe. We discuss some implications of our results in \S\ref{sec:discussion}.

\section{Numerical calculation}

\subsection{Simulation setup}\label{sec:numerical_setup}

We carry out a 3D relativistic hydrodynamic simulation of a jet launched into an expanding stellar envelope, using the code \textsc{pluto} \citep{Mignone2007}. We follow the jet propagation in the ejecta, its activity after breakout, and the period after it shuts off. We post-process the simulation output semi-analytically to produce estimates of the post-breakout emission. As the simulations are dimensionless, we can calibrate the output with dimensionless mass scale $ m_s $ and timescale (length scale) $ t_s $.

The jet operates for $ \teng = 4460\,t_s\,\s $ with a total (two-sided) luminosity, $ \Lj = 2\times 10^{47}\,m_s/t_s\,{\rm erg\,\s^{-1}} $. We consider as fiducial values $\{t_s = 200, m_s = 3\}$, corresponding to jet duration $\teng = 10.3\,\days $, and luminosity $L_{\rm j} = 3\times 10^{45}\,{\rm erg}\,\s^{-1} $, typical of the engine properties needed to power Type I SLSNe (Fig.~\ref{fig:engineproperties}; see also \citealt{Metzger+15}, their Fig.~1). These jet properties are illustrated as an orange square in Fig.~\ref{fig:engineproperties}, while bidirectional arrows indicate the change in engine properties obtained by varying $t_s$ ($L_{\rm eng} \propto t_{\rm eng}^{-1}$) that would result in the same hydrodynamic evolution (for fixed ejecta properties $m_s$). This degeneracy running approximately parallel to the SLSN population suggests that the hydrodynamic evolution predicted by our single simulation could apply to a wide range of SLSN engines. For the early bump to be observable, the jet breakout must occur early enough to avoid being outshone by the primary SN emission. Eq.~\ref{eq:tbo} demonstrates that, for typical jet and ejecta properties, the breakout timescale is indeed expected to be significantly shorter than the peak of the main SLSN.

Building on the results of \citet{Gottlieb2022d}, we set the jets at the inner boundary to be precessing about the spin axis of the black hole, with an inclination angle of $ 0.05 $ rad and a period of $ t_s\,\s $. The initial jet Lorentz factor is $ \Gamma_0 = 10 $ with a maximum Lorentz factor\footnote{We note that the precise ultra-relativistic Lorentz factor of the jet does not affect the cocoon, which maintains $ \Gamma \lesssim 3 $ at all times \citep{Gottlieb2021a}.} of 20 (equipartition between initial thermal and kinetic energy), and its opening angle is $ \theta_0 = 0.1 $ rad. The jet runs into a homologously expanding $ M_{\rm ej} = 7\,m_s\msun $ ejecta with a front velocity of $ \vej = 10^4\,\km\,\s^{-1} $ and initial position $ \reji = 7\times 10^{6}\,t_s\,\km $. The ejecta initial comoving mass density profile is:

\begin{equation}
    \rho(r) = \left(\frac{r}{9\times 10^{5}\,\km}\right)^{-1.5}\left(1-\frac{r}{\rej}\right)^3\frac{m_s}{t_s^3}\frac{\rm g}{\rm cm^3}\,.
\end{equation}
	
We use an ideal gas equation of state with a polytropic index of 4/3, as appropriate for radiation-dominated gas. This applies at all relevant times for our calculation as we focus on the outflow while radiation is coupled to the gas. The 3D Cartesian grid is divided into three patches on the $ \hat{x} $- and $ \hat{y} $-axes: The inner patch is at the innermost $ 2\times 10^{5}\,t_s\,\km $ with 100 uniformly distributed cells. The outer patches are stretched from $ \lvert 2\times 10^{5}\,t_s\,\km\rvert $ to $ \lvert 3\times 10^8\,t_s\,\km\rvert $ with 350 logarithmically distributed cells. We divide the $ \hat{z} $-axis (jet axis) into four patches: starting from $ z_0 = 2\times 10^5\,t_s\,\km $ to $ 3\times 10^6\,t_s\,\km $ with 300 uniformly distributed cells, followed by three logarithmically spaced patches with 600 cells to $ 3\times 10^7\,t_s\,\km $, 350 cells to $ 9.5\times 10^7\,t_s\,\km $, and 480 cells to $ 6\times 10^8\,t_s\,\km $. The jet is injected at the lower boundary, $ z_0 $, with a cylindrical injection radius of $ r_0 = z_0\theta_0 $. The integration is performed with a Harten-Lax-van Leer (HLL) solver, Runge-Kutta time stepping, and piecewise parabolic interpolation.

\subsection{Emission calculation}\label{sec:numerical_emission}

We first consider the released thermal near-ultraviolet (NUV)/optical emission from the gas that escaped the trapping radius in the last dynamical time. We assume that each fluid element is propagating radially at the relevant times of emission, such that we calculate the trapping radius along each line of sight $ \Omega $, and time $ t $:
\begin{equation}\label{eq:trapping}
\rtr(t,\Omega)=R\left[\tau(t,\rtr,\Omega)=\frac{1}{\beta(t,\rtr,\Omega)}\right]\,,
\end{equation}
where $ \beta $ is the dimensionless velocity, and $ \tau $ is the optical depth along a radial line of sight:
\begin{equation}\label{eq:tau}
    \tau(t,r,\Omega) = \int_{r}^{\infty}\kappa\rho(t,r',\Omega)\left[1-\beta\left(t,r',\Omega\right)\right]\gamma(t,r',\Omega)\mathrm{d}r'\,,
\end{equation}
where $ \kappa = 0.1\,\cm^2\,{\rm g}^{-1} $ is the opacity of the gas (e.g., \citealt{Kleiser&Kasen18}), and $ \gamma $ is its Lorentz factor.  

For two given lab times, $ t_1, t_2 $, we find the trapping radius in the Lagrangian frame, $ \rtr^{L1}, \rtr^{L2} $. We calculate the thermal energy between these radii as:
\begin{equation}
    \Eth\left(\bar{t},\Omega\right) = \int_{\rtr^{L2}}^{\rtr^{L1}}{3p(\bar{t},r,\Omega)\gamma(\bar{t},r,\Omega)r^2 dr}\,,
\end{equation}
where $ p $ is the comoving gas thermal pressure and $ \bar{t} = \frac{t_1+t_2}{2} $. As the emitting gas is sub-relativistic with $ \vemit \equiv \beta(\rtr) \ll 1 $, we neglect relativistic effects for the luminosity and the relative arrival times of the photons. The thermal energy is released during $ t_2-t_1$, such that the local luminosity in the lab frame is:
\begin{equation}
    L(\bar{t},\Omega) = \frac{\Eth\left(\bar{t},\Omega\right)}{t_2-t_1}\,,
\end{equation}
and the bolometric luminosity is:
\begin{equation}
    L_{\rm bol}(\bar{t}) = \int L(\bar{t},\Omega)d\Omega\,,
\end{equation}
where the integration is performed over one hemisphere, as the diffusion time through the equator exceeds the dynamical time ($ \tau \beta > 1 $), restricting observers near the axis to observing only one jet.

Eq.~\eqref{eq:trapping} implies that the optical depth of the non-relativistic emitting shell is always much greater than unity. Thus, the diffusing photons will interact with free electrons before reaching the photosphere. These interactions will govern the photon energy at the photosphere, defined as $ \rph(t,\Omega)=R\left[\tau(t,\rph,\Omega)=1\right] $. At the relevant times and temperatures, there is enough time for the gas to produce photons through free-free emission to share the photon energy and reduce the temperature from the trapping radius until reaching thermalization \citep[see discussion in][]{Nakar2010}. Therefore, the local temperature at the photosphere is taken to be a blackbody:
\begin{equation}
    T(\bar{t},\Omega) = \left[\frac{L\left(\bar{t},\Omega\right)}{4\pi\sigma_b \rph\left(\bar{t},\Omega\right)^2}\right]^{1/4}\,.
\end{equation}
Using the local temperature, we calculate the local spectral luminosity, $ L_\nu $, at various bands, and integrate it over all angles to obtain the observed spectral luminosity. From the spectral luminosity, we infer the color temperature as:
\begin{equation}
    T_c(\bar{t}) = \frac{h\nu_{\rm max}(\bar{t})}{2.821\,k_B}\,,
\end{equation}
where $ h $ and $ k_B $ are the Planck and Boltzmann constants, respectively, and $ \nu_{\rm max}(\bar{t}) = {\rm arg\,max}\left[L_\nu(\bar{t})\right]$.

Finally, as the jet interacts with the high-density CSM, it powers afterglow synchrotron emission. We follow the semi-analytic solution in the thin-shell approximation of \citet{Oren2004} for the propagation of a top-hat jet in a wind medium, to calculate the resulting radio emission.

\section{Emission}\label{sec:results}

\subsection{NUV/Optical}

    \begin{figure}
        \centering
        \includegraphics[width=3.3in]{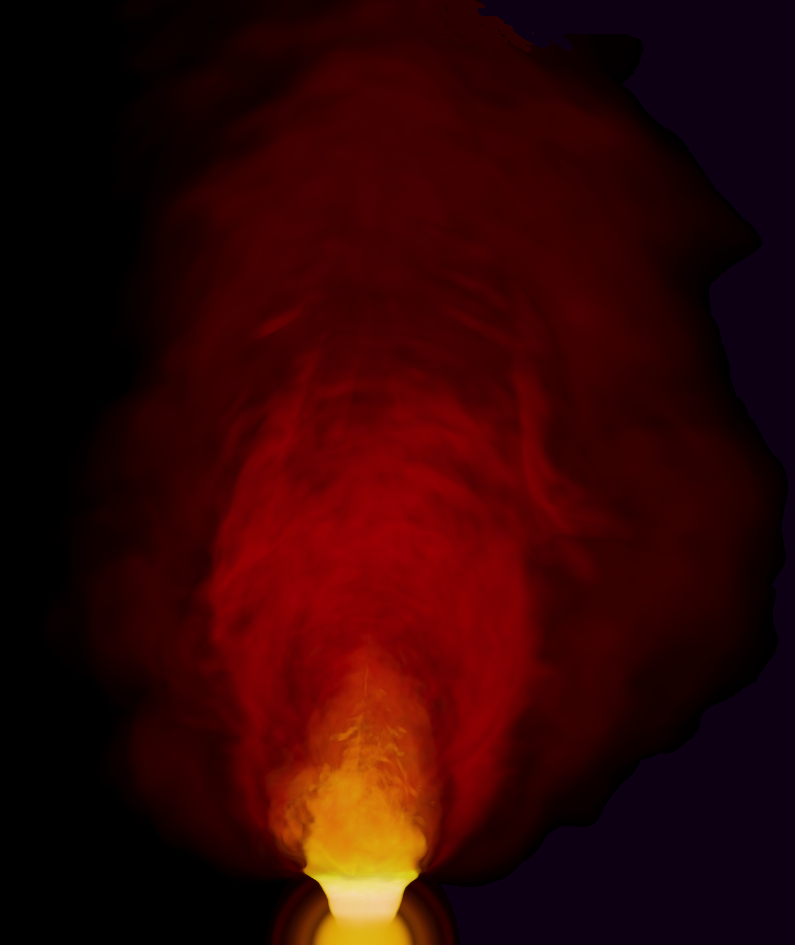}
        \caption{
        3D rendering of the logarithmic comoving mass density at the end of the jet activity, $ t = \teng $, when the ejecta front is at $ \rej \approx 8.5\times 10^{14}\,\km $. As the jet escapes from the high-density expanding spherical ejecta (yellow), it generates a stratified structure of the lower-density cocoon (red). The interaction of the jet and the cocoon with the expanding ejecta close to the breakout radius generates bright NUV/Optical emission over a $ \sim $ week timescale.
        }
        \label{fig:3D}
    \end{figure}

    \begin{figure*}
        \centering
        {\includegraphics[width=3.5in,trim={0cm 0cm 0cm 0cm}]{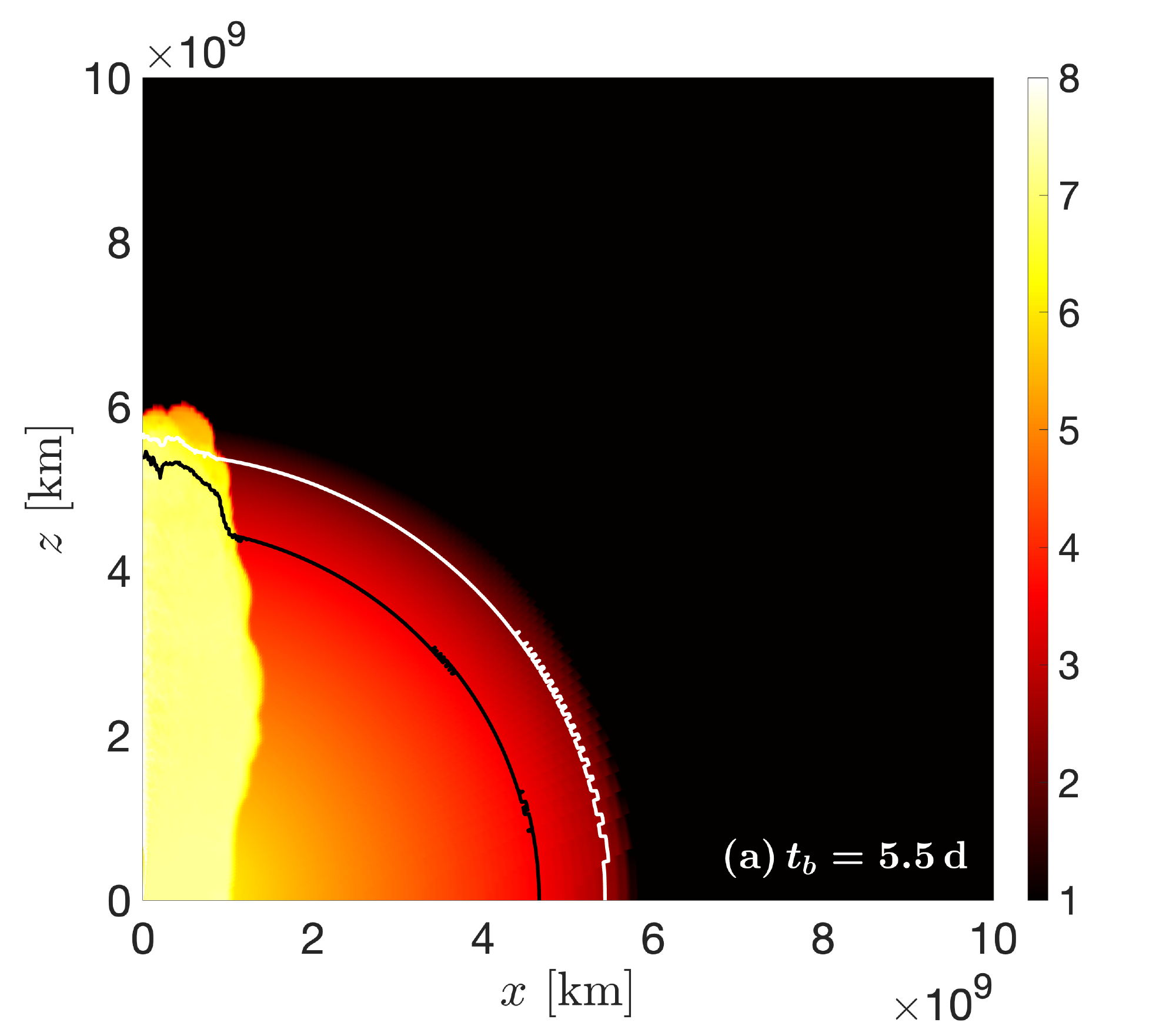}}
        {\includegraphics[width=3.5in,trim={0cm 0cm 0cm 0cm}]{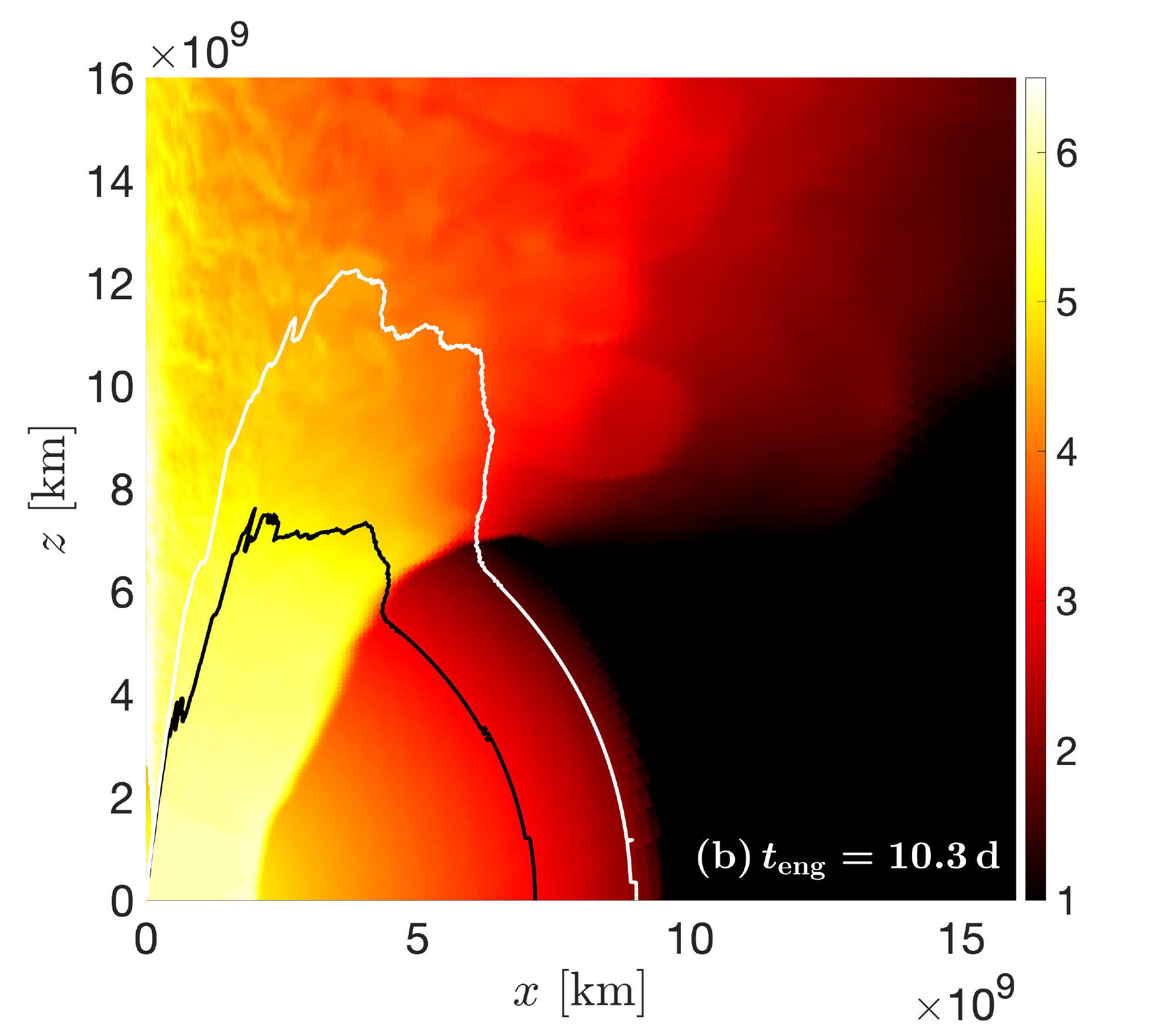}}
        {\includegraphics[width=3.5in,trim={0cm 0cm 0cm 0cm}]{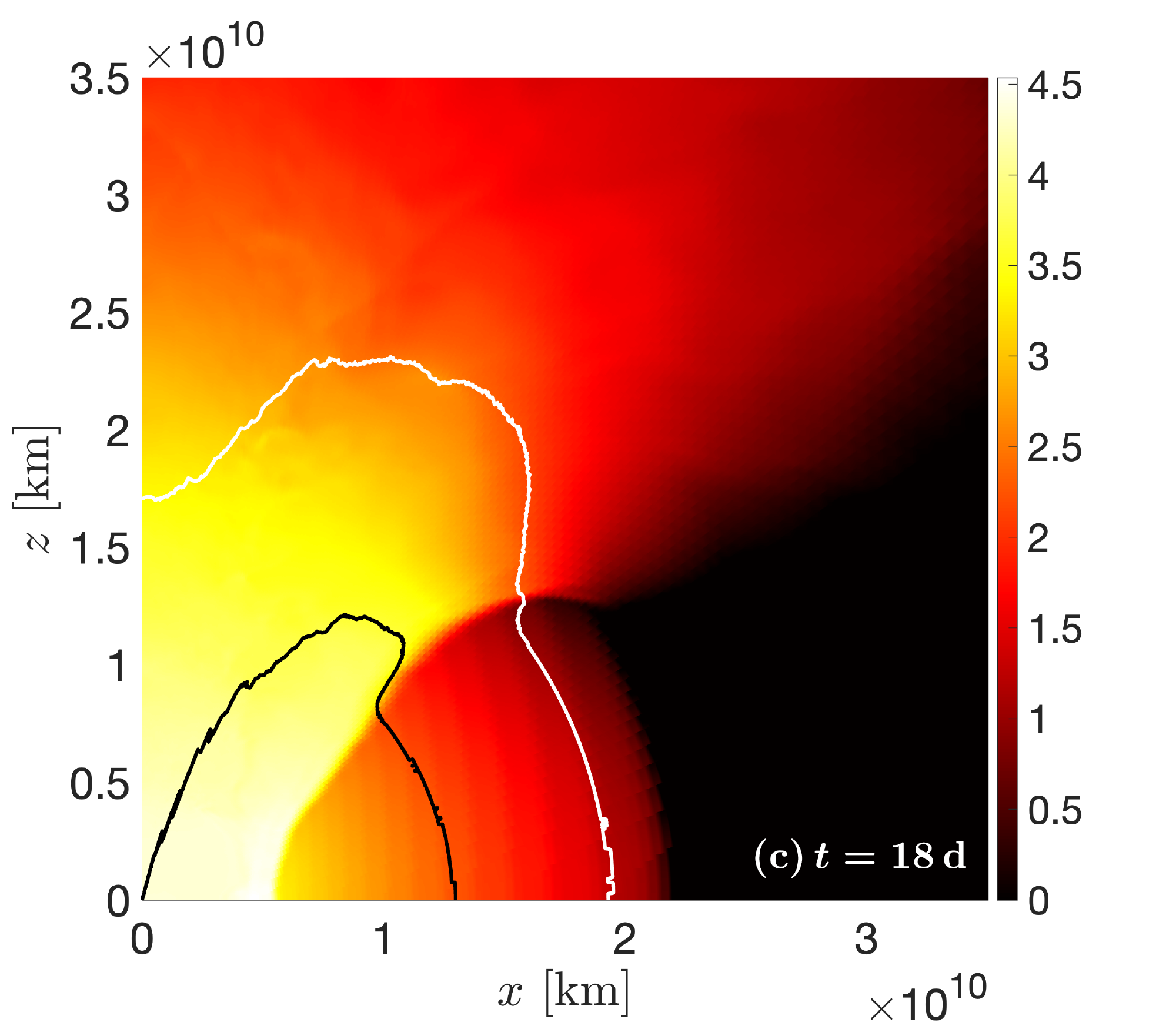}}
        {\includegraphics[width=3.5in,trim={0cm 0cm 0cm 0cm}]{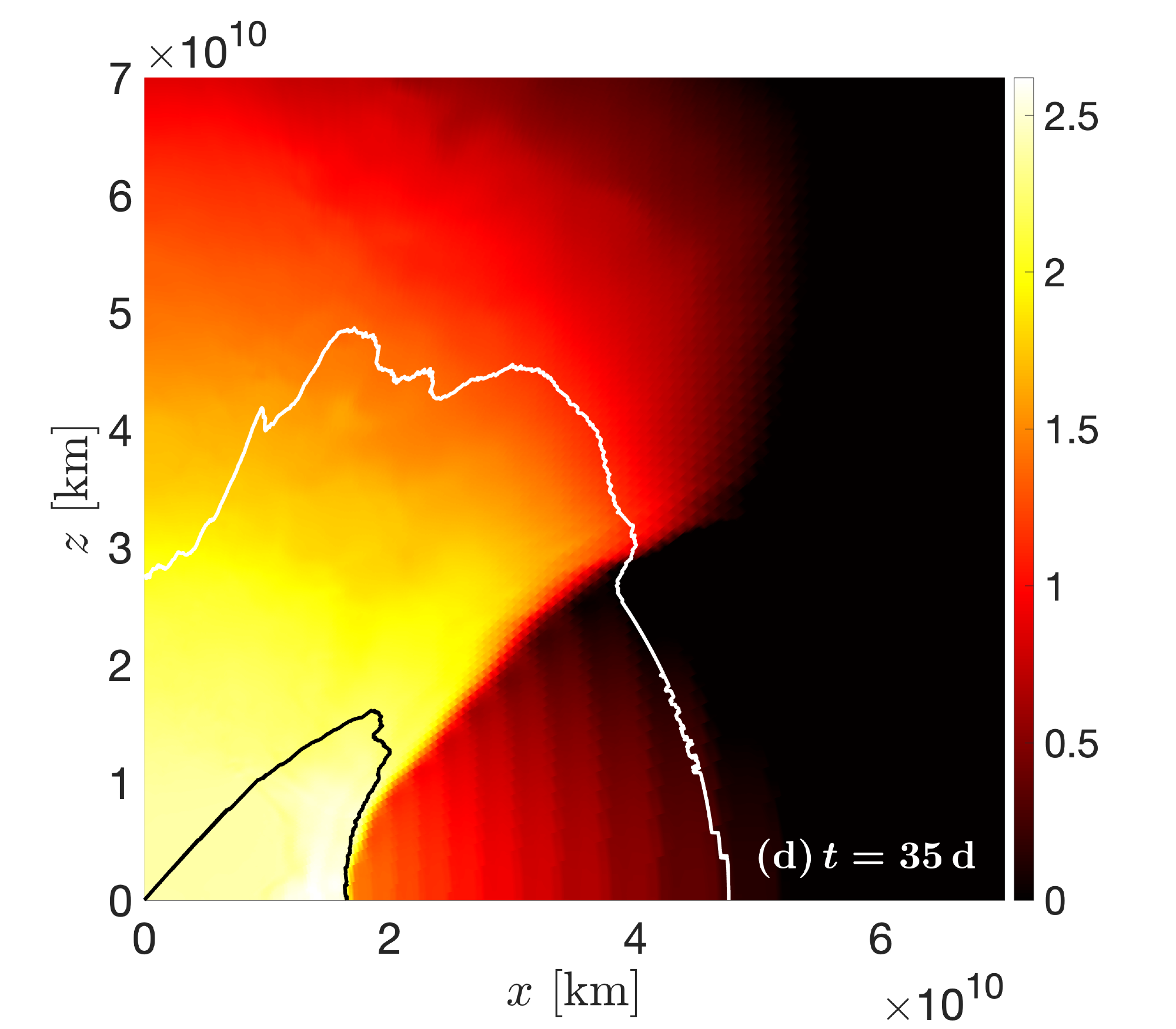}}
        \caption{2D $ \hat{x}-\hat{z} $-plane cuts at various times of the azimuthally-averaged logarithmic comoving thermal energy density, $ e = 3p $ maps. The white and black contours represent the photosphere and trapping radius, respectively.
        }
        \label{fig:maps}
    \end{figure*}

    \begin{figure*}
        \centering
        {\includegraphics[width=3.5in,trim={0cm 0cm 0cm 0cm}]{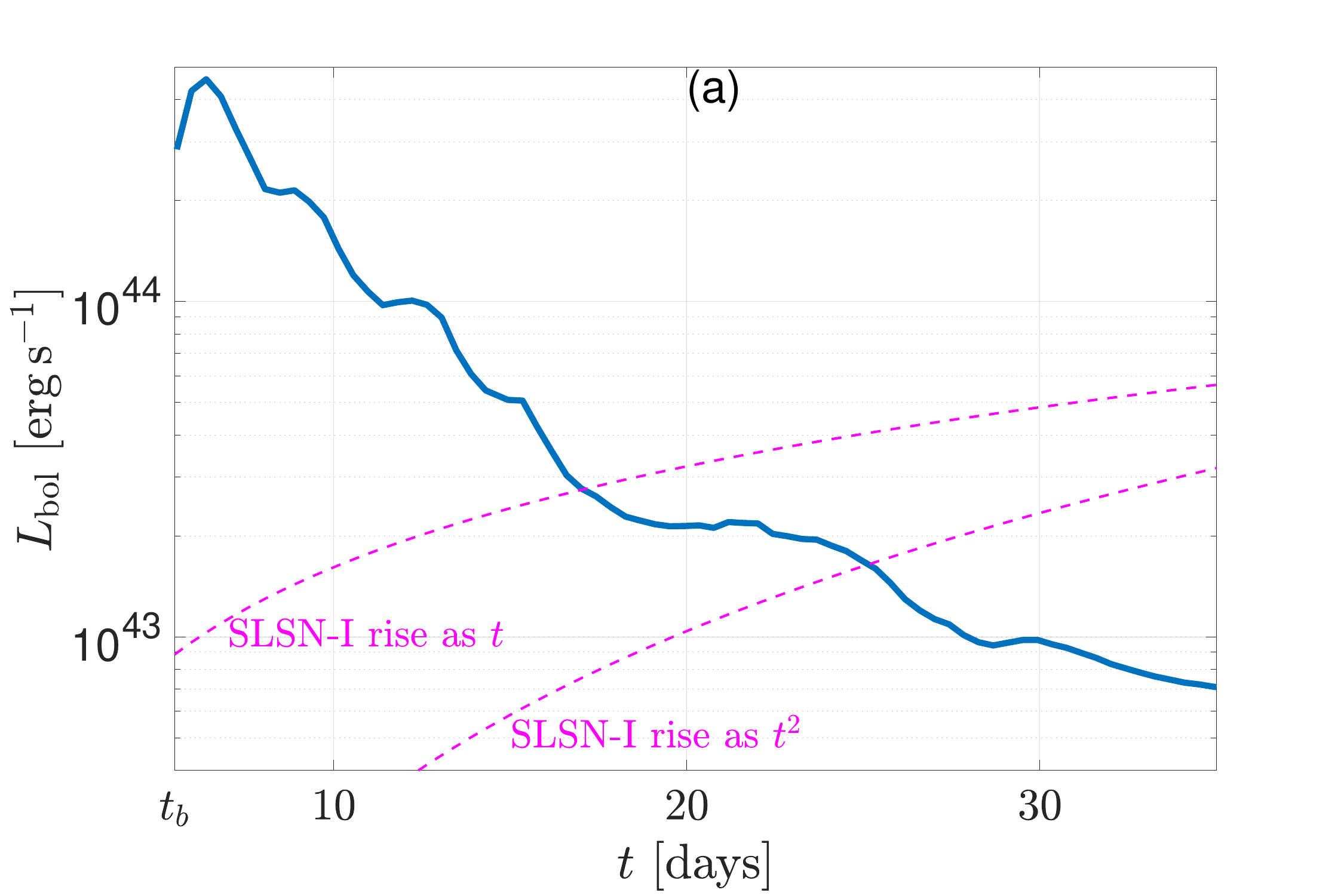}}
        {\includegraphics[width=3.5in,trim={0cm 0cm 0cm 0cm}]{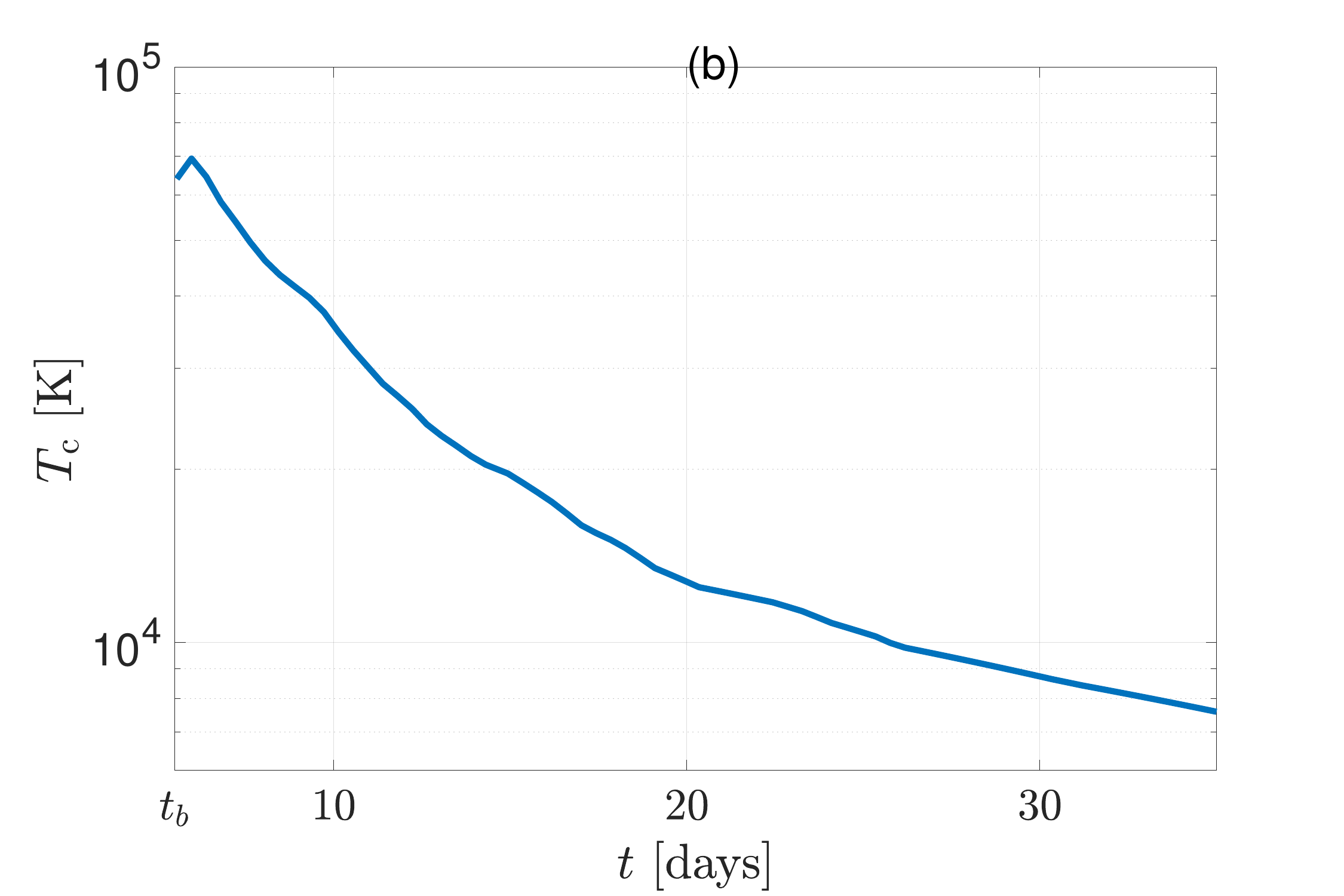}}
        {\includegraphics[width=3.5in,trim={0cm 0cm 0cm 0cm}]{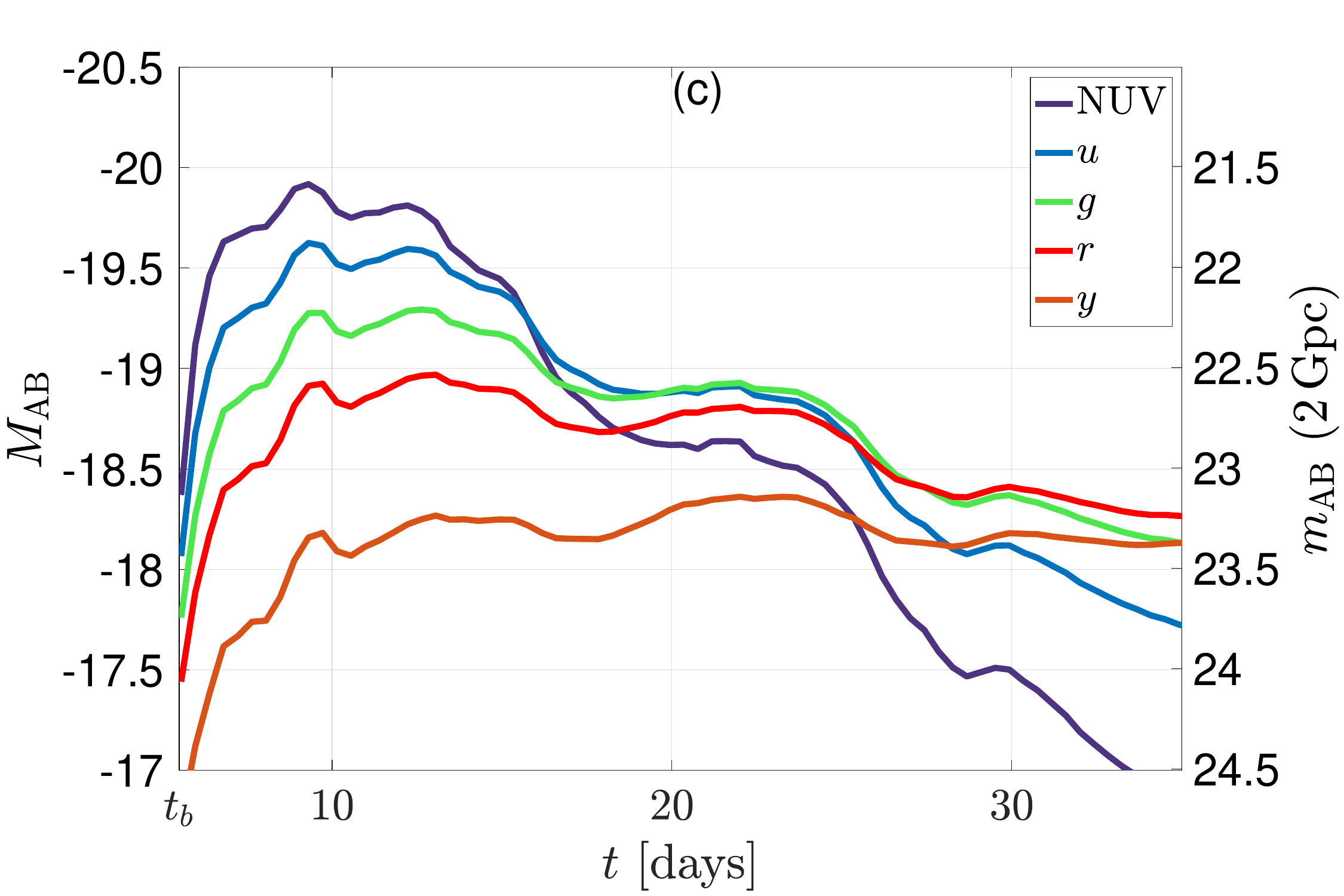}}
        {\includegraphics[width=3.5in,trim={0cm 0cm 0cm 0cm}]{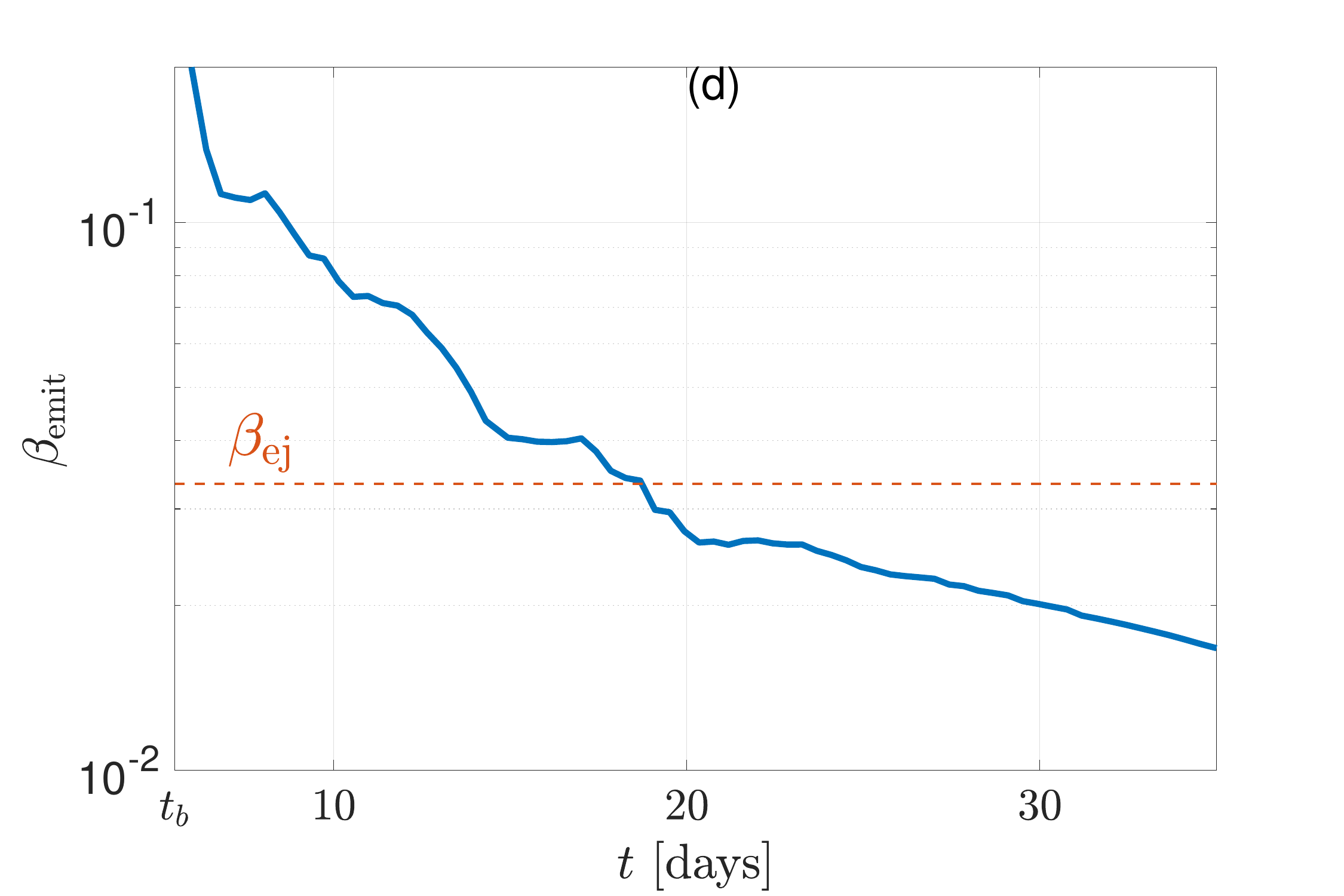}}
        \caption{
        {\bf (a)}: Bolometric luminosity peaks soon after breakout ($ t_b $), and then declines. The dashed magenta lines illustrate the rise of the main Type I SLSN peak.
        {\bf (b)}: Color temperature of the radiating gas.
        {\bf (c)}: Absolute (left vertical axis) and apparent (right vertical axis) magnitude at various bands ($ f_{\rm NUV} = 1.2\times 10^{15}\,{\rm Hz} $, $ f_{u} = 8.6\times 10^{14}\,{\rm Hz} $, $ f_{g} = 6.3\times 10^{14}\,{\rm Hz} $, $ f_{r} = 4.8\times 10^{14}\,{\rm Hz} $, $ f_{y} = 3.0\times 10^{14}\,{\rm Hz} $) demonstrate that the peak magnitude aligns with observations of SLSN early bumps.
        {\bf (d)}: Luminosity-weighted average dimensionless velocity of the radiating gas at the trapping radius demonstrates that the emitting gas is always subrelativistic. The dimensionless front ejecta velocity is shown in red for comparison.
        }
        \label{fig:lightcurve}
    \end{figure*}

The head of the weak jet propagates slowly inside the expanding ejecta with an average velocity in the ejecta frame of $ v_h-\vej \approx 0.01c $. This implies that all the jet energy is practically deposited into the cocoon for as long as the jet head remains within the ejecta. For our choice of parameters: $ t_s = 200 $ and $ m_s = 3 $, the two-sided jet with a luminosity of $ \Lj = 3\times 10^{45}\,\erg\,\s^{-1} $ breaks out from the $ M_{\rm ej} \approx 20\,\msun $ ejecta after $ t_b \approx 5.5\,\days $, indicating that the cocoon energy upon breakout is $ L_c \approx \Lj t_b \approx 1.7\times 10^{51}\,\erg $. Figure~\ref{fig:3D} illustrates a 3D rendering of the logarithmic comoving mass density, showcasing the interaction of the jet-cocoon outflow (red) with the homologously expanding ejecta (yellow). This interaction generates emission near the optically thin breakout radius and shapes the large-scale stratified structure observed beyond the confined expanding ejecta.

To elucidate the different phases of the evolution and emission, Figure~\ref{fig:maps} presents azimuthally-averaged logarithmic comoving thermal energy density maps at various times, where the black and white contours delineate the trapping radius and photosphere, respectively. Figure~\ref{fig:lightcurve} displays the resultant $ L_{\rm bol} $ (a), $ T_c $ (b), absolute and apparent magnitudes (c), and $ \vemit $ (d). using the numerical calculation described in \S\ref{sec:numerical_emission}. We identify three phases of emission:

(1) The first emission phase begins upon breakout and continues for as long as the jet engine is active at $ t_b < t < t_{\rm eng} $. Fig.~\ref{fig:maps}(a) depicts the breakout of the collimated jet and its hot, pressurized cocoon from the expanding ejecta. Upon breakout, photons above the trapping radius escape from the top layers of the cocoon, producing the first light through shock breakout emission. At this time, the emitting layers are mildly relativistic, as indicated by the proximity of the trapping radius to the photosphere. Shortly after the shock breakout signal, the gas from deeper layers starts to contribute to the emission with $ \vemit \approx 0.2 $ [Fig.~\ref{fig:lightcurve}(d)], and the luminosity begins to fall off [Fig.~\ref{fig:lightcurve}(a)]. The high temperature immediately after breakout of $ T_c \approx 7\times 10^4\,{\rm K} $ [Fig.~\ref{fig:lightcurve}(b)] results in dominant ultraviolet emission [Fig.~\ref{fig:lightcurve}(c)].

After the breakout, the jet head accelerates to ultra-relativistic velocities, and the mildly relativistic cocoon expands in the angular direction to $ \theta \sim (\Gamma\beta)^{-1} $. In typical GRBs, the breakout takes place within $ \sim 100\,\s $, implying that the photons generated in the jet-cocoon interaction are still trapped well within the outflow. Here, Figure~\ref{fig:maps}(b), which depicts the jet-cocoon system at the time of jet engine shutoff at $ t_{\rm eng} = 10.3\,\days $, suggests that the jet is optically thin, and the late breakout results in a trapping radius located just outside of the jet-cocoon interface \citep[JCI;][]{Gottlieb2021a} and within the ejecta outer radius. Consequently, the emission originates from the continuous heating in the JCI, as indicated by the high thermal energy along the increasing trapping radius in the angular direction away from the jet axis. Over time, the drop in the temperature [Fig.~\ref{fig:lightcurve}(b)] results in a transition to NUV emission which peaks at $ t \approx t_{\rm eng} $ [Fig.~\ref{fig:lightcurve}(c)].

(2) Figures~\ref{fig:maps}(c,d) demonstrate that after the jet shuts off, the polar cavity is loaded with baryon and turns optically thick. At the same time, the emission from the jet-cocoon interaction ceases, resulting in a drop in the bolometric luminosity at $ t \approx t_{\rm eng} $. At this point, the thermal energy along the angular direction of the trapping radius becomes roughly homogeneous. The dominant heat source then becomes the newly shocked material from the lateral spreading of the cocoon into the unshocked ejecta below the trapping radius, with $ \vemit \lesssim \beta_{\rm ej} $, where $ \beta_{\rm ej} $ is the dimensionless front ejecta velocity [Fig.~\ref{fig:lightcurve}(d)]. As the heat source is located below the trapping radius, the emission follows a cooling-envelope model, with the light curve evolving as $ L \sim t^{-2} $, as expected from the expansion of the sub-relativistic gas in the cocoon \citep{Gottlieb2022b}.

The high temperatures at jet activity cause the emission to first peak in the ultraviolet, followed by a peak in the optical bands after the jet engine is shut off. As the luminosity declines more rapidly than the temperature at $ t \gtrsim t_{\rm eng} $, the near-infrared $ y $ -band emission rises since it is located on the Rayleigh-Jeans portion of the spectrum. Consequently, the decrease in the bolometric luminosity following jet shutdown also reduces the optical emission. 

(3) Comparing the panels in Fig.~\ref{fig:maps}, it is evident that the trapping radius shrinks in the ejecta frame, and the entire unshocked ejecta has become optically thin in Fig.~\ref{fig:maps}(d). The final phase of the emission will occur once the cocoon becomes optically thin. The timing of this transition depends on the opacity, which may decrease with temperature, depending on the composition of the SN ejecta. In particular, if the ejecta is primarily composed of O-Ne-Mg nuclei, the electron scattering and line-opacity drops substantially once $ T \approx 6\times 10^3\,{\rm K} $, at $ t \approx 5\,\days $ due to recombination effects (e.g., \citealt{Kleiser&Kasen18}). However, if the line opacity of Fe-group nuclei (e.g., Fe-Co-Ni) dominates, the opacity will remain relatively constant across this temperature range, causing the optically thin transition to occur somewhat later.

Ultimately, the rise of the main SN emission, powered primarily by a (more isotropic, non-jet-related) source of ejecta heating by the same central engine (e.g., the absorption of X-rays and gamma-rays from the magnetar nebula; \citealt{Vurm&Metzger21}), will take over the optical light curve. The dashed magenta curves in the bolometric luminosity panel illustrate the rise of the Type I SLSN emission towards the main peak at $ L_{\rm bol} \approx 10^{44}\,\erg\,\s^{-1} $ at $ t \approx 50\,\days $ \citep[e.g.,][]{Quimby+11,GalYam12}. We present both a slow rise, $ L_{\rm bol} \sim t $, and a fast rise, $ L_{\rm bol} \sim t^2 $. 

Our light curve calculation considers the integrated emission over all solid angles. However, an observer will only detect the whole emission within certain viewing angles relative to the jet axis, where most of the thermal energy is generated. The first light -- shock breakout emission [Figure~\ref{fig:maps}(a)] is quasi-isotropic with mild beaming due to the mildly relativistic velocities of the front cocoon. After the initial signal, the emitting gas becomes subrelativistic [Figure~\ref{fig:lightcurve}(d)], so it is no longer subject to relativistic beaming effects. However, the emitted photons can only reach observers along optically thin lines of sight, i.e. ``geometric'' beaming may still play a role. Figure~\ref{fig:maps}(b) shows that the phase in which the jet is active is only observable within the JCI at $ \theta_{\rm obs} \lesssim 2\theta_j $. Over time, as the cocoon expands angularly, the accessible angle gradually increases. Therefore, the light curve observed at angles farther from the jet axis may peak at shock breakout, decay, and rise again at later times compared to the bolometric luminosity shown in Figure~\ref{fig:lightcurve}(a). Full radiation transfer calculations are needed to provide a self-consistent modeling of the light curves at different viewing angles.

\subsection{Radio}

Ultimately, the jet will decelerate due to its interaction with the dense CSM, powering an ``orphan'' radio afterglow synchrotron emission and an optical flash from the reverse shock \citep{Sari1999}. However, the latter is likely to be outshone by the thermal cocoon emission associated with the early bump. While the lack of orphan afterglow detections in blind surveys \citep[][see however \citealt{Perley2024} for a recent potential detection]{Ho2018}, along with the potential for high CSM densities around some Type I SLSNe \citep[e.g.,][]{Margutti+23} that would amplify the signal, may argue against the generic presence of jets in these systems, the large uncertainties prevent us from reaching definitive constraints.

    \begin{figure}
        \centering
        {\includegraphics[width=3.6in,trim={0cm 0cm 0cm 0cm}]{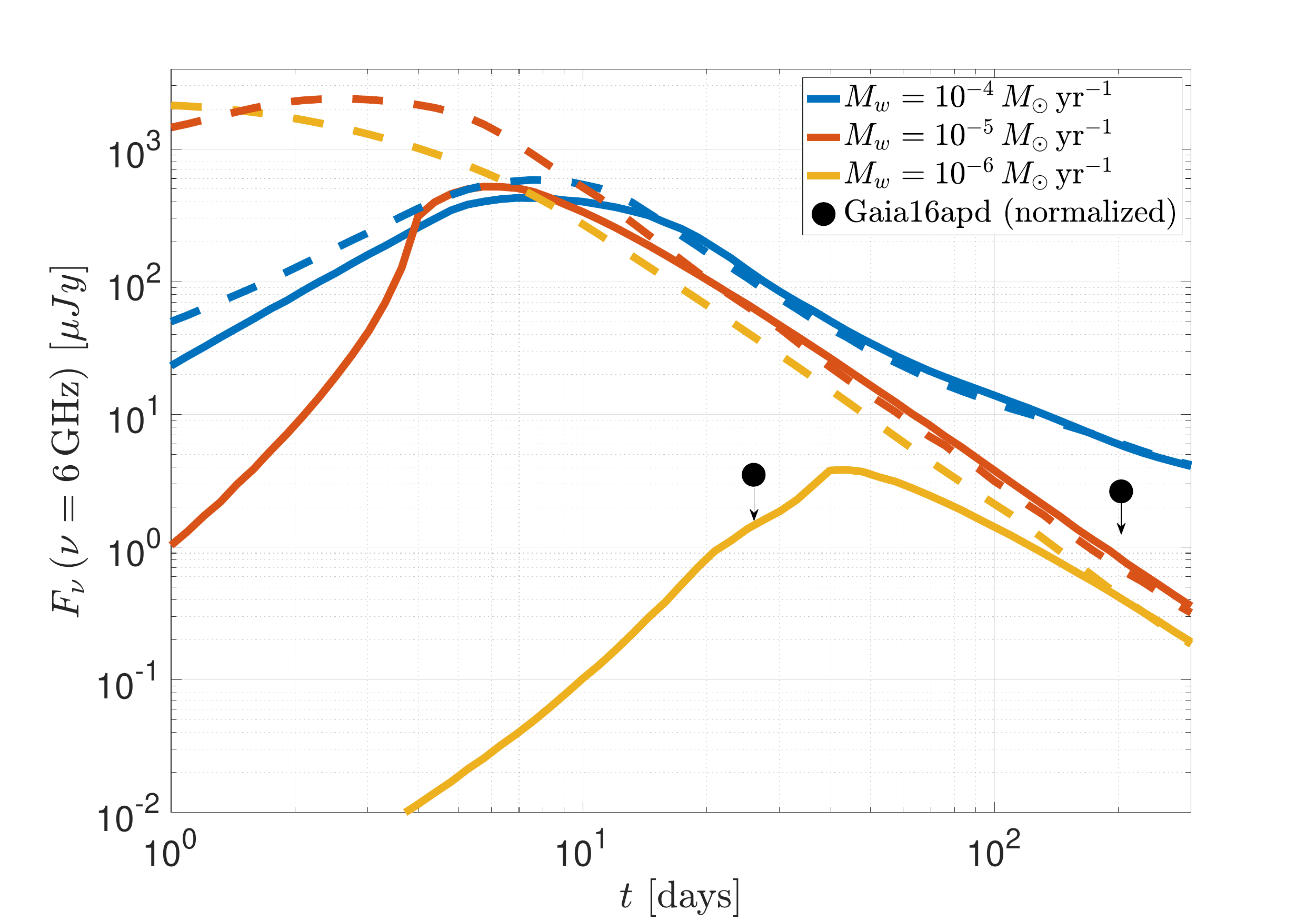}}
        \caption{
        Top-hat jet afterglow light curves at $ \nu = 6\,{\rm GHz} $ for observers at $ \theta_{\rm obs} = 0.5 \,{\rm rad} $ (solid lines) and on-axis observers (dashed lines) at $ z = 0.25 $, assuming jet Lorentz factor $ \Gamma = 100 $, opening angle $ \theta_j = 0.1\,{\rm rad} $, and the following microphysics parameters: electron equipartition parameter $ \epsilon_e = 0.1 $, magnetic equipartition parameter $ \epsilon_B = 10^{-4} $, electron energy power-law index $ p = 2.2 $, and wind velocity $ v_w = 1000\,\km\,\s^{-1} $. The black circles denote the most constraining upper limits measured for Type I SLSN Gaia2016apd \citep{Coppejans+18}, placing the source at $ z = 0.25 $.
        }
        \label{fig:afterglow}
    \end{figure}

Figure~\ref{fig:afterglow} presents the semi-analytic thin-shell approximation solution for the radio synchrotron emission of our model at various assumed progenitor wind mass-loss rates, for both off-axis (solid lines) and on-axis (dashed lines) viewing angles. We assume a characteristic redshift of $ z = 0.25 $ for Type I SLSNe \citep[e.g.,][]{Gomez+24}, standard jet properties with $ \Gamma = 100 $ and $ \theta_j = 0.1\,{\rm rad} $, and microphysics parameters (detailed in the caption) inferred from GW170817 afterglow and very long baseline interferometry images \citep{Mooley2018}. If the mass-loss rates are $ M_w \gtrsim 10^{-5}\,\msun\,{\rm yr}^{-1} $, the emission is self-absorbed during the first $ \sim 1 $ week, leading to an initial rise in the light curves for all observers. 

The extended engine activity indicates that the jet's initial radial extent is $ \sim (t_{\rm eng} - t_b)c \sim 10^{16}\,\cm $. This suggests that not all of the jet's energy is transferred to the forward shock before the jet's width becomes comparable to the shock width, $ \sim R/12\,\Gamma^2 $, at an observed time of $ \sim 12\,(t_{\rm eng}-t_b) \sim 1 $ month. This duration could be shorter if, for example, the jet's radial extent shrinks beforehand due to high CSM densities. As a result, the light curve, calculated under the thin-shell approximation, may overestimate the radio emission during this initial period, thus providing an upper limit.

Beyond the general uncertainties in the afterglow and jet parameters, the strong dependence of the flux on the viewing angle \citep{Nakar2002} and the wind mass-loss rate complicates robust predictions that can support or refute our model. The most stringent upper limits on Type I SLSNe were obtained for Gaia2016apd \citep{Coppejans+18} at $ z = 0.1018 $ \citep{Yan2017}. The black circles represent these upper limits (adjusted for $ z = 0.25 $), potentially ruling out some scenarios with high mass-loss rates and small viewing angles for this source, but not excluding off-axis observers and lower mass-loss rates within the considered parameters.

\section{Discussion}\label{sec:discussion}

Cocoon emission following the birth of a compact object has been invoked in several energetic transients, including collapsars (e.g., \citealt{Nakar2015,Nakar2017}) and neutron star mergers (e.g., \citealt{Lazzati2017,Gottlieb2018a,Gottlieb2018b}). However, in general, the cocoon emission in these systems is radiatively inefficient, such that a significant fraction of the deposited jet energy goes into adiabatic losses rather than observable radiation. By contrast, the longer duration of the engines at work in SLSNe, which inject the bulk of their energy over days to weeks instead of seconds to minutes, and the correspondingly much lower densities of the SN ejecta at these late times, reduce the impact of adiabatic losses. In particular, the trapping radius moves inside the JCI on a timescale comparable to the engine duration. This enables an order-unity fraction of the jet luminosity to emerge as radiation.

Such a high radiative efficiency is important because the luminosities of the early bumps seen in SLSN light curves $\sim 10^{43}-10^{45}$ erg s$^{-1}$ correspond to a sizable fraction $\sim 1-100\%$ of the total energy budget of the engine (Fig.~\ref{fig:engineproperties}). In the case of a magnetar engine, the fraction of the magnetar spin-down energy placed into a jet depends on several uncertain conditions, such as the inclination angle of the magnetic dipole (\citealt{Margalit+18b}, which determines the fraction of the striped pulsar wind that undergoes magnetic reconnection at the wind termination shock; e.g., \citealt{Lyubarsky03,Komissarov13}) and the efficiency with which any remaining large-scale field avoids instabilities and associated reconnection \citep{Bucciantini+09,Porth+13,Mosta2015}. 
In the case of a black hole engine, the power of the relativistic jet depends on the magnetic field and spin of the black hole \citep{Blandford&Znajek77}, while accretion disk outflows can release their energy more isotropically and contribute to powering the bulk of the SLSN emission (e.g., \citealt{Dexter&Kasen13}). If the jet engine operates for $ \teng \lesssim t_b $, then the jet will be choked deep inside the ejecta, resulting in no powerful breakout and hence absent or weaker associated UV/optical emission. A substantial failed jet fraction is compatible with early maxima being detected in only 3/14 Type I SLSNe observed by DES \citep{Angus+19}.

For several days after the jet emerges from the star, the cocoon emission is hot, $T_{\rm c} \gtrsim 3\times 10^{4}$ K, and peaks in the ultraviolet at an absolute magnitude $M_{\rm AB} \approx -20$. Although multi-band measurements of early SLSN bumps which constrain their temperatures are sparse, such data on PS1-10pm indeed indicated high $\approx (2-3)\times 10^{4}$ K temperatures during its early bump phase \citep{Nicholl&Smartt16}. Similar temperature evolution was observed from the early maximum in DES14x3taz, which cooled from $\approx 2.5\times 10^{4}$ K at discovery to $\lesssim 10^{4}$ K over roughly a week \citep{Smith+16}.

A few wide-field UV satellites with time-domain capabilities are planned over the next decade, including the Ultraviolet Transient Astronomy Satellite (ULTRASAT, launch date 2027; \citealt{Sagiv+14}), Ultraviolet Explorer (UVEX; \citealt{Kulkarni+21}), and the Czech mission QUVIK \citep{Werner+23}. We focus on ULTRASAT, which will reach a $5\sigma$ sensitivity of 22.4 AB magnitude in the NUV band ($h \nu \approx $ 5 eV) across an instantaneous field of view of $\approx 200$ deg$^{2}$ for a 900 second (15 minutes) integration \citep{Sagiv+14}.\footnote{
UVEX will possess sensitivity extending also into the FUV and reaching an AB magnitude depth of 24.5 for a 900 s integration \citep{Kulkarni+21}, roughly a factor of 6 deeper than ULTRASAT. However, UVEX's smaller instantaneous field of view $\approx 12$ deg$^{2}$, results in a comparable survey speed time to ULTRASAT.} Most of the observing time will be spent on a low cadence survey, cycling through $\Delta \Omega \sim 6800$ deg$^{2}$ ($f_{\Omega} = 0.16$ of the whole sky) covering 10 fields of view per day (4-day cadence per field). 

For a source with $M_{\rm AB} \approx -20$ lasting several days, we estimate an ULTRASAT horizon distance of $D_{\rm lim} \approx 3$ Gpc. We assume the early NUV emission to be detectable over $ \theta_{\rm obs} \approx 0.5\,{\rm rad} $. Given the volumetric rate of Type I SLSNe of $\mathcal{R}\approx 40$ Gpc$^{-3}$ yr$^{-1}$ \citep{Quimby+13,Frohmaier+21,Zhao+21} at $z \sim 0.2-1$, and assuming that all SLSNe produce early jet breakout emission similar to what we have predicted, the number of detectable sources per year can be estimated as $N_{\rm det} = (2\pi/3)\theta_{\rm obs}^2 D_{\rm lim}^{3}\mathcal{R}f_{\Omega} \sim 10^{2}\,{\rm yr}^{-1} $. The detection rates will be lower if jets are less powerful than our baseline model, or only successful in a fraction of explosions, as discussed above and supported by Type I SLSNe observations \citep{Angus+19}.  More direct rate estimates could be made based on a large uniform sample of premaximum SLSN light curve observations \citep{Nicholl&Smartt16}. Given its greater depth and similar cadence, many of the ULTRASAT sources will also have optical light curve data from the {\it Vera C. Rubin Observatory} \citep{Ivezic+19}, enabling better constraints on the temperature evolution.  

An early photometric detection could also enable prompt spectra to be obtained.  We predict that the earliest phases of the emission could reveal Doppler-broadened velocities extending to $v/c \gtrsim 0.1$ (bottom right panel of Fig.~\ref{fig:lightcurve}), at least for observers relatively close to the jet axis.  These and other jet breakout signatures, such as luminous X-ray emission for face-on observers (possibly already observed in one SLSN; \citealt{Levan+13}) or radio afterglow emission from the jet or its cocoon (\citealt{Coppejans+18,Eftekhari+21}; Fig.~\ref{fig:afterglow}), would provide the ultimate ``smoking gun'' of an engine-powered origin for SLSNe.  
Our model for powering the early bumps in SLSNe bears some resemblance to the proposal by \citet{Kasen+16}, insofar that we are also invoking shock heating of the SN ejecta due to energy input from a central engine which is substantially delayed after the explosion. However, while \citet{Kasen+16} consider emission from a spherical (or quasi-spherical) shock breakout driven by the high pressure of a nebula (``bubble'') inflated by a central engine (see also \citealt{Suzuki&Maeda16}), we have considered emission from the interaction of a tightly collimated jet with its cocoon, closer to the proposal of \citet{Margalit+18b}. Furthermore, for the nebula-driven shock to reach the ejecta surface, the energy of the central engine must exceed the original kinetic energy of the explosion \citep{Kasen+16}, while collimated jet breakout from homologous ejecta places a less stringent requirement on the engine energy (e.g., \citealt{Gottlieb&Nakar22}). The successful breakout of a relativistic jet also comes with additional consequences, such as a synchrotron afterglow from relativistic ejecta, that in general will not be present in the spherical breakout case (however, see \citealt{Arons03,Bucciantini+11}). The outer layers of the ejecta accelerated by a spherical breakout can produce their own synchrotron emission (e.g., \citealt{Maeda&Suzuki23}), but exhibit a different light curve and spectral evolution than a jet afterglow, particularly for observers close to the jet axis.  

The direct detection of internal high-energy radiation from the jet viewed on-axis would provide an even more dramatic confirmation of the model. The mechanism, radiative efficiency, and spectrum of any internal jet emission are highly uncertain. Nevertheless, we focus on the soft X-ray band, motivated by the X-ray detection of SCP 06F6 \citep{Levan+13} and the recent launch of the satellite mission {\it Einstein Probe}, which reaches a sensitivity in the soft X-ray band across a large fraction of the sky of $F_{\rm lim} \approx 10^{-12}$ erg cm$^{-2}$ s$^{-1}$ for a $10^{5}$ s integration \citep{Yuan+22}, comparable to the expected duration of SLSN jets. Assuming that a fraction $f_{\rm X}$ of the jet power $L_{\rm j}$ is placed into soft X-rays, the resulting detection horizon is $D_{\rm lim} \approx (f_{\rm X}L_{\rm j}/4\pi F_{\rm lim})^{1/2}$. Adopting a beaming fraction $f_{\rm b} \sim 10^{-2}$ similar to those of GRB jets, the number of detectable on-axis SLSN jets per year can be estimated as:
\begin{equation}
N_{\rm X} \approx \frac{4\pi}{3}D_{\rm lim}^{3}\mathcal{R}f_{\rm b} \sim 1.3\,{\rm yr^{-1}}\left(\frac{f_{\rm X}}{0.1}\right)^{3/2}\left(\frac{L_{\rm X}}{10^{45}\,{\rm erg\,s^{-1}}}\right)^{3/2}
\end{equation}
It is thus possible that Einstein Probe or successor missions could detect on-axis Type I SLSN jets over the coming years, although given the many uncertainties, a lack of such detections is not necessarily constraining.

\acknowledgements

We thank Sebastian Gomez and Matt Nicholl for sharing data from their recent compilation of SLSN light curves and for helpful feedback on the manuscript. We also thank the referee, Ben Margalit, Raffaella Margutti, and Dan Kasen for useful comments. OG is supported by the Flatiron Research Fellowship. BDM was supported in part by the National Science Foundation (grant No. AST-2009255) and by the NASA Fermi Guest Investigator Program (grant No.~80NSSC22K1574). The Flatiron Institute is supported by the Simons Foundation. The computations in this work were, in part, run at facilities supported by the Scientific Computing Core at the Flatiron Institute, a division of the Simons Foundation.

\bibliography{refs,refsBDM}

\end{document}